\documentclass[twocolumn,aps]{revtex4} 

\def\XXint#1#2#3{{\setbox0=\hbox{$#1{#2#3}{\int}$}
     \vcenter{\hbox{$#2#3$}}\kern-.5\wd0}}

\usepackage{epsfig}
\begin{document} 
%\draft 
\title{Analytical solution of the Gross-Neveu model at finite density}  
\author{Michael Thies\footnote{Electronic address: thies@theorie3.physik.uni-erlangen.de}} 
\address{Institut f\"ur Theoretische Physik III\\ 
Universit\"at Erlangen-N\"urnberg\\ 
Staudtstra\ss e 7\\ 
D-91058 Erlangen\\ 
Germany} 
\date{\today} 
\begin{abstract}
Recent numerical calculations have shown that the ground state of the Gross-Neveu model
at finite density is a crystal. Guided by these results, we can now present the analytical solution
to this problem in terms of elliptic functions. The scalar potential is the superpotential of
the non-relativistic Lam\'{e} Hamiltonian. This model can also serve as analytically
solvable toy model for a relativistic superconductor in the Larkin-Ovchinnikov-Fulde-Ferrell 
phase. 
\end{abstract} 

%string pacs\{\}} should always be input, 
%even if empty.} 
%\narrowtext 
\maketitle 

In this paper we reconsider the simplest variant of the Gross-Neveu (GN) model,
a 1+1 dimensional relativistic field theory with $N$ species of fermions
interacting via a quartic self-interaction \cite{GrossNeveu}
\begin{equation}
{\cal L} = \bar{\psi}^{(i)} {\rm i} \gamma^{\mu} \partial_{\mu} \psi^{(i)}
+ \frac{1}{2} g^2 \left( \bar{\psi}^{(i)}\psi^{(i)}\right)^2 \, .
\label{A1}
\end{equation}
In a previous work \cite{ThiesUrlichs}, we have found that 
the widely accepted phase diagram of this model in the large $N$ limit \cite{Wolff}
needed some revision.
The dynamically generated scalar mean field becomes inhomogeneous in a certain
region of temperature and chemical potential, a fact which had been overlooked so far.
The four-fermion interaction then does not merely lead to mass generation
but to formation of a kink-antikink crystal. This in turn reflects the presence of bound baryons
in the GN model as can be most clearly seen in the low-density limit. 
The approach used in \cite{ThiesUrlichs} was a numerical implementation of the Dirac-Hartree-Fock
method (equivalent to the saddle point method in the functional integral approach). Although
one can in principle carry out such calculations to any desired accuracy, it remains 
somewhat embarrassing to have to rely on numerics when dealing with a supposedly
exactly solvable model.

In the meantime we have been able to overcome this deficiency.
In the present work, we focus on the $T=0$ case and explain how to construct the
crystal ground state at any density in closed, analytical form. We expect that the finite
temperature calculation can be done along similar lines but have to leave this for future work. 
Since the results of Ref. \cite{ThiesUrlichs} are fully confirmed by our new method
we refer to this paper for more details, figures and a discussion
of the underlying physics. 
For a more general introduction into the field of 1+1 dimensional toy models for hot and
dense matter, see the review article \cite{SchoenThies}.
Here we concentrate on the technical details of the analytical solution.

We start from the Hartree-Fock Dirac equation,
\begin{equation}
\left( \gamma^5 \frac{1}{\rm i} \frac{\partial}{\partial x} + \gamma^0 S(x)\right)
\psi(x) = \omega \psi(x) \, ,
\label{1}
\end{equation}
choosing the $\gamma$-matrices as follows,
\begin{equation}
\gamma^0 =-\sigma_1 \, , \quad \gamma^1 ={\rm i}\sigma_3 \, , \quad \gamma^5 = \gamma^0
\gamma^1 =-\sigma_2 \, .
\label{2}
\end{equation}
In terms of the upper and lower spinor components $\phi_{\pm}$ 
the Dirac equation consists of two coupled equations 
\begin{eqnarray}
\left( \frac{\partial}{\partial x} - S \right) \phi_- &=& \omega \phi_+ \nonumber \\
- \left( \frac{\partial}{\partial x} + S \right)\phi_+ & = & \omega \phi_- \, .
\label{4}
\end{eqnarray}
which can be decoupled by squaring,
\begin{equation}
\left( - \frac{\partial^2}{\partial x^2} \mp \frac{\partial S}{\partial x} + S^2 \right)
\phi_{\pm} = \omega^2 \phi_{\pm} \, .
\label{5}
\end{equation}
Note that Eqs.~(\ref{4},\ref{5}) fall precisely into the pattern of supersymmetric (SUSY) quantum
mechanics. Let us now make an ansatz for $S(x)$ based on the  
superpotential of the well-known Lam\'{e} potential \cite{WhittakerWatson,Feinberg} 
\begin{equation}
S(x)=A \kappa^2 \frac{{\rm sn}(Ax| \kappa^2) {\rm cn}(Ax|\kappa^2)}
{{\rm dn}(Ax| \kappa^2)} \equiv A \tilde{S}(Ax) \, .
\label{6}
\end{equation}
Here, three types of Jacobi elliptic functions with modulus $\kappa$ appear
\cite{Jacobi}. 
Denoting the rescaled space coordinate $Ax$ by $\xi$, Eq.~(\ref{5}) becomes
\begin{equation}
\left( - \frac{\partial^2}{\partial \xi^2} \mp \frac{\partial \tilde{S}}{\partial \xi}+\tilde{S}^2
\right) \phi_{\pm} = \frac{\omega^2}{A^2} \phi_{\pm} \, .
\label{7}
\end{equation}
The spatial period of $\tilde{S}(\xi)$ is 
\begin{equation}
\ell = 2 \mathbf{K}
\label{8}
\end{equation}
where $\mathbf{K}$ is the complete elliptic integral of the first kind, $\mathbf{K}(\kappa^2)$ \cite{Jacobi}.
We shall choose the parameter $A$ in such a way that the original potential $S(x)$
has the period $a$ determined by the mean density \cite{ThiesUrlichs},
\begin{equation}
a=\frac{1}{\rho}=\frac{\pi}{p_f} \, ,
\label{9}
\end{equation}
hence
\begin{equation}
A=\frac{\ell}{a} = \frac{2 p_f \mathbf{K}}{\pi}\, .
\label{10}
\end{equation}
The resulting potential still has one free parameter, $\kappa$, which determines both its shape and
its size; the period is now fixed by the mean density. Equation (\ref{7}) for $\phi_+$ can then be converted 
into
\begin{equation}
\left( - \frac{\partial^2}{\partial \xi^2} + 2 \kappa^2 {\rm sn}^2(\xi | \kappa^2) \right)
\phi_+ = {\cal E} \phi_+
\label{11}
\end{equation}
with
\begin{equation}
{\cal E} = \frac{a^2}{\ell^2} \omega^2 + \kappa^2 \, ,
\label{12}
\end{equation}
which is recognized as the simplest case of the Lam\'{e} equation \cite{WhittakerWatson}. 
The corresponding equation for $\phi_-$ differs only
by a translation of the potential through half a period and thus yields an identical spectrum \cite{Feinberg}. 
Our particular ansatz for the scalar potential was of course designed in such a way 
as to map the Dirac equation onto a soluble Schr\"odinger equation with a periodic potential.
We can now simply use all the well-known results for the Lam\'{e} potential. For our purpose we 
found Ref. \cite{Lietal} particularly useful.

In order to determine the yet unknown parameter $\kappa$ we shall minimize the ground state energy
density. In the Hartree-Fock approach, this is usually decomposed as
\begin{equation}
E_{\rm g.s.}= - 2 \int_{p_f}^{\Lambda/2} \frac{{\rm d}p}{2\pi} \omega
+ \frac{1}{2Ng^2a}\int_0^a {\rm d}x S^2(x) \equiv E_1 + E_2.
\label{13}
\end{equation}
$E_1$ is the sum over single particle energies over all filled negative energy states
regularized by a cutoff.
(For simplicity, we consider antimatter by leaving the valence band in the Dirac sea
unoccupied \cite{ThiesUrlichs}.) $E_2$ is the standard correction term for double counting of the
interaction energy. Consider $E_1$ first, transforming the Bloch momenta $p$ and
single particle  energies $\omega$
to the corresponding quantities from the Lam\'{e} equation, 
\begin{equation}
E_1 = -2 \frac{\ell^2}{a^2} \int_{k_{\min}}^{k_{\max}} \frac{{\rm d}k}{2\pi} \sqrt{{\cal E}-\kappa^2}
\label{14}
\end{equation}
where 
\begin{equation}
k_{\min}= \frac{\pi}{\ell} \, , \quad k_{\max} = \frac{a\Lambda}{2\ell}\, .
\label{15}
\end{equation}
It is actually more convenient to integrate over ${\cal E}$, using \cite{Lietal}
\begin{equation}
\frac{{\rm d}k}{{\rm d}{\cal E}}= \frac{
\mathbf{E}/\mathbf{K}+ \kappa^2-{\cal E}}{2 \sqrt{(1-{\cal E})
({\cal E}-\kappa^2)(1+\kappa^2-{\cal E})}} \, .
\label{16}
\end{equation}
Here, $\mathbf{E}$ is the complete elliptic integral of the second kind, $\mathbf{E}(\kappa^2)$.
We thus have to evaluate
\begin{equation}
E_1 = - 2 \frac{\ell^2}{a^2}\int_{{\cal E}_{\min}}^{{\cal E}_{\max}} \frac{{\rm d}{\cal E}}
{2\pi}  \left| \frac{{\rm d}k}{{\rm d}{\cal E}}\right|  \sqrt{{\cal E}-\kappa^2}   
\label{17}
\end{equation}
where now the lower limit is the band edge,
\begin{equation}
{\cal E}_{\min} = 1 + \kappa^2\, ,
\label{18}
\end{equation}
whereas the upper limit can be inferred from Eq.~(\ref{16}) to be
\begin{equation}
{\cal E}_{\max} =k_{\max}^2 + 2 \left(1- \frac{\mathbf{E}}{\mathbf{K}} \right) 
\label{19}
\end{equation}
with $k_{\max}$ given in Eq.~(\ref{15}). It is necessary to keep the sub-leading
term here, since the integral over ${\rm d}{\cal E}$ is linearly divergent and the divergent
part will be subtracted. Performing the integration in Eq.~(\ref{17}) yields
\begin{eqnarray}
E_1 &=& -\frac{\Lambda^2}{8\pi} + \frac{\ell^2}{4\pi \mathbf{K} a^2} \left(4\mathbf{E}+(\kappa^2-2)\mathbf{K}\right)
\nonumber \\
& & +\frac{\ell^2}{2\pi \mathbf{K} a^2}\left( 2\mathbf{E} + (\kappa^2-2)\mathbf{K}\right) \ln \left( \frac{a \Lambda}
{\ell \kappa} \right) \, .
\label{20}
\end{eqnarray}
The term $\sim \Lambda^2$ can be eliminated by subtracting the energy of the trivial
vacuum. Now consider the double counting correction $E_2$, Eq.~(\ref{13}), 
in the form
\begin{equation}
E_2 = \frac{\ell}{2 Ng^2 a^2} \int_0^{\ell} {\rm d}\xi \tilde{S}^2(\xi) \, .
\label{21}
\end{equation}
Inserting $\tilde{S}$ from Eq.~(\ref{6})  and transforming to the integration variable $s={\rm sn}\, \xi$,
the integration can be carried out as follows,
\begin{eqnarray}
\int_0^{\ell} {\rm d}\xi \frac{{\rm sn}^2 \xi \, {\rm cn}^2 \xi}{{\rm dn}^2 \xi}
& = &  2 \int_0^1 {\rm d}s  \frac{s^2 \sqrt{1-s^2}}{(1-\kappa^2s^2)^{3/2}} \nonumber \\
& = &  -\frac{2}{\kappa^4} \left( 2 \mathbf{E} + (\kappa^2-2)\mathbf{K} \right) \, .
\label{22}
\end{eqnarray}
The coupling constant is related to the cutoff via the (vacuum) gap equation
\cite{ThiesUrlichs,Zeitschrift} which reads 
(in units where the vacuum fermion mass is 1)
\begin{equation}
N g^2 = \frac{\pi}{\ln \Lambda} \, .
\label{23}
\end{equation}
Combining Eqs.~(\ref{21})--(\ref{23}), we find
\begin{equation}
E_2 = - \frac{\ell}{\pi a^2} \left( 2 \mathbf{E} + (\kappa^2-2)\mathbf{K}\right) \ln \Lambda \, .
\label{24}
\end{equation}
Upon adding $E_1$ and $E_2$ and recalling Eq.~(\ref{8}), the logarithmically divergent terms cancel and 
we obtain the finite, renormalized ground state energy density,
\begin{eqnarray}
E_{\rm ren} &=&  \frac{p_f^2 \mathbf{K}}{\pi^3}\left( 4 \mathbf{E} + (\kappa^2-2)\mathbf{K}\right)
 \\
& & + \frac{2 p_f^2 \mathbf{K}}{\pi^3} \left(2\mathbf{E}
+(\kappa^2-2)\mathbf{K}\right) \ln \left(
\frac{\pi}{2 p_f  \kappa \mathbf{K}}\right)\, . 
\nonumber
\label{25}
\end{eqnarray}
Let us minimize this expression with respect to the modulus $\kappa$, our variational parameter. 
This yields the simple condition
\begin{equation}
\kappa = \frac{a}{\ell} = \frac{\pi}{2 p_f \mathbf{K}} \, ,
\label{26}
\end{equation}
a transcendental equation for $\kappa$. Eliminating 
 $p_f$ from $E_{\rm ren}$ with the help of this relation, we finally get the following 
parametric representation of the ground state energy as a function of density
(parameter $\kappa$),
\begin{eqnarray}
E_{\rm ren} &=& \frac{1}{4\pi} + \frac{1}{\pi \kappa^2} \left( \frac{\mathbf{E}}{\mathbf{K}}-\frac{1}{2}\right)\, ,
\label{26a1}
 \\
\frac{p_f}{\pi} &=& \frac{1}{2\kappa \mathbf{K}}\, .
\label{26a2}
\end{eqnarray}
We also give the scalar potential $S(x)$ corresponding
to the optimal value of the modulus $\kappa$, 
\begin{equation}
S(x)= \kappa \frac{{\rm sn}\left(\frac{x}{\kappa}\Big| \kappa^2\right)
 {\rm cn}\left(\frac{x}{\kappa} \Big|\kappa^2\right)}{{\rm dn}\left(\frac{x}{\kappa}\Big| \kappa^2\right)}\, .
\label{26b}
\end{equation}
$S(x)$
interpolates smoothly between widely spaced kinks and antikinks ($\sim \pm \tanh x$) at $\kappa \to 1$
(low-density limit)
and the function $\frac{\kappa}{2} \sin \left( \frac{2 x}{\kappa}\right)$ for $\kappa \to 0$
(high-density limit).
Unfortunately, it does not seem possible to express $E_{\rm ren}$ or $S(x)$ directly in
terms of $p_f$ since the relation between $p_f$ and $\kappa$,
Eq.~(\ref{26a2}), cannot be analytically inverted. 
 
Let us pause here and summarize what has been achieved so far. All we have done may be
viewed as a variational calculation of the ground state of baryonic matter in the
GN model. Our variational ansatz amounts to generating single particle orbits from a
scalar potential $S(x)$ and filling all occupied negative energy levels. Guided by
severe restrictions from analytical solvability, we choose the one-parameter family of scalar potentials
defined in Eqs.~(\ref{6}) and (\ref{10}).  The result of varying the parameter $\kappa$
is given in Eqs.~(\ref{26a1})--(\ref{26b}). This in itself does not sound very exciting.
However, if we now compare the present results with those of Ref. \cite{ThiesUrlichs},
we discover that the analytical variational solution thus obtained
agrees perfectly with the solution of the numerical Hartree-Fock
calculation. At all densities considered in \cite{ThiesUrlichs},   
both the values of the ground state energy and the shape and depth of the scalar potentials are 
indistinguishable if plotted in one graph. In the numerical calculation,
no bias about the shape of the potential was put in (except for periodicity), since
the Fourier components of $S(x)$ were used as independent variational parameters.
We therefore conclude that the true ground state happens to lie on the one-parameter
trajectory of potentials which can be dealt with analytically. In order to show this
without invoking any numerical results, one still has to verify that the  
ground state expectation value of $\bar{\psi}\psi$ is self-consistent, as was done for the
single baryon in \cite{Zeitschrift}. This is indeed possible, but requires more details
about the rather involved Lam\'{e} wave functions \cite{WhittakerWatson}
as well as some patience in juggling identities for elliptic functions. In order 
to keep this paper readable, we have therefore deferred the full analytic proof to the appendix.

Let us now make use of the closed formulae derived above to illustrate certain features of the
GN crystal. If  we go to the low- or high-density limit, 
it becomes possible to systematically resolve the
transcendental equation relating $p_f$ and $\kappa$,
\begin{eqnarray}
\kappa  &  \begin{array}[t]{c} \approx \\
\raisebox{1ex} {\mbox{${\scriptstyle p_f \to \, 0}$}}
\end{array}  & 1 - 8 {\rm e}^{-\pi/p_f} +\frac{32(\pi + p_f)}{p_f}{\rm e}^{-2\pi/p_f} 
\nonumber \\
\kappa  &  \begin{array}[t]{c} \approx \\
\raisebox{1ex} {\mbox{${\scriptstyle p_f \to \, \infty}$}}
\end{array} &
\frac{1}{p_f} - \frac{1}{4 p_f^3} + \frac{3}{64 p_f^5} 
\label{27x1}
\end{eqnarray}
The non-analytic dependence of $\kappa$ on $p_f$ at low density is a reflection of tunnelling 
between the widely separated baryon wells.
For the energy as a function of density, one finds 
\begin{eqnarray}
E_{\rm ren}&  \begin{array}[t]{c} \approx \\
\raisebox{1ex} {\mbox{${\scriptstyle p_f \to \, 0}$}}
\end{array}  &  -\frac{1}{4\pi} + \frac{2 p_f}{\pi^2} + \frac{8 p_f}{\pi^2} {\rm e}^{-\pi/p_f} 
\nonumber \\
E_{\rm ren}  &  \begin{array}[t]{c} \approx \\
\raisebox{1ex} {\mbox{${\scriptstyle p_f \to \, \infty}$}}
\end{array} & \frac{p_f^2}{2\pi} - \frac{1}{2^6 \pi p_f^2} + \frac{3}{2^{14}\pi p_f^6} 
\label{27x2}
\end{eqnarray}
In the low-density limit, the three terms correspond to the vacuum energy density, the 
contribution from the baryon mass ($\sim \rho M_B$ with $M_B=2/{\pi}$) and a term describing
the repulsive baryon-baryon interaction. At high densities, we can identify the free massless
Fermi gas piece, the leading perturbative correction already given in \cite{ThiesUrlichs} and the next term
coming from higher order effects, suggesting fast convergence. 
In view of the comparison with \cite{ThiesUrlichs} it is also instructive to determine the Fourier coefficients
$S_n$ of $S(x)$, 
\begin{equation}
S(x)=\sum_n S_n {\rm e}^{{\rm i}2\pi n x/a} \ , \qquad (S_{-n}=S_n^*) \ .
\label{27y}
\end{equation}
Upon using Eq.~(16) on p. 912 of Gradshteyn-Ryzhik \cite{Gradshteyn} and correcting
a misprint ($\pi^2$ should read $\pi$ on the right hand side), we find the following closed expression
(only odd $n$'s appear),
\begin{equation}
{\rm i}S_n =  \frac{2 p_f}{\sinh (n \pi \mathbf{K'}/\mathbf{K})} \, ,
\label{27a}
\end{equation}
where 
\begin{equation}
\mathbf{K'} = \mathbf{K}(1-\kappa^2) \, . 
\label{27b}
\end{equation}
Low- and high-density limits of our previous work \cite{ThiesUrlichs} can now easily be confirmed, 
namely
\begin{eqnarray}
{\rm i}S_n  &  \begin{array}[t]{c} \approx \\
\raisebox{1ex} {\mbox{${\scriptstyle p_f \to \, 0}$}}
\end{array}  &   \frac{2 p_f}{\sinh (n \pi p_f)}
\nonumber \\
{\rm i}S_1  &  \begin{array}[t]{c} \approx \\
\raisebox{1ex} {\mbox{${\scriptstyle p_f \to \, \infty}$}}
\end{array} & \frac{1}{4 p_f}
\label{27c}
\end{eqnarray}
and again one finds excellent agreement with the numerical results at all densities.

Finally, we wish to point out that the GN model at finite density can also serve as
a solvable model for a relativistic, inhomogeneous superconductor. Along the lines
described in Ref. \cite{Thies}, one can map the GN Lagrangian onto a ``dual" Lagrangian
which has quark-quark rather than quark-anti-quark pairing. All one has to do is redefine
quarks into anti-quarks for left-handed quarks only. If one works at non-zero chemical potential,
a baryonic chemical potential $\mu$ in the GN model corresponds to an ``axial" chemical potential
$\mu_5$ in the dual BCS-type model. Left-handed and right-handed fermions have opposite chemical potentials,
hence $\mu_5$ in 1+1 dimensions acts like a magnetic field in 3+1 dimensions. This favors 
the appearance of the Larkin-Ovchinnikov-Fulde-Ferrell 
(LOFF) phase with spatially varying Cooper pair condensate \cite{LO,FF}. It is then natural
to identify the kink-antikink crystal of the GN model with the LOFF phase of the dual BCS-type model. In this
sense, the present study may also be of some use for model studies of relativistic superconductors. 

\begin{center}
{\bf Appendix}
\end{center}
{\bf Analytical proof of self-consistency of $S(x)$}
\\
We would like to show that
\begin{equation}
S(x)=-Ng^2 \sum_{\alpha}^{\rm occ} \bar{\psi}_{\alpha}(x) \psi_{\alpha}(x) \, ,
\label{s1}
\end{equation}
where the sum runs over all negative energy levels correponding to the ``upper band" 
of the Lam\'{e} spectrum (we are again considering antimatter).
First, we have to construct normalized spinor solutions of the Dirac equation (\ref{1})
out of the known solutions of the Lam\'{e} equation (\ref{11}). 
We write
\begin{equation}
\psi = {\cal N} \left( \begin{array}{c} \phi_+ \\ \phi_- \end{array} \right)
\label{s3}
\end{equation}
and choose for $\phi_+$ a solution of the 2nd order differential equation (\ref{11}). We recall that the
relations between Dirac variables and Lam\'{e} variables are
\begin{equation}
{\cal E}= \kappa^2(\omega^2+1) \ , \qquad x=\kappa \xi \ , \qquad p = k /\kappa \ , 
\label{s4}
\end{equation}
where we have used condition (\ref{26}). 
Once $\phi_+$ is chosen, $\phi_-$ follows from Eq.~(\ref{4}), 
\begin{equation}
\phi_- = -\frac{1}{\kappa \omega} \left( \frac{\partial}{\partial \xi} + \tilde{S}(\xi)\right)
\phi_+ \ . 
\label{s5}
\end{equation}
Let us first compute the normalization factor ${\cal N}$. In a continuum normalization, the
(spatially averaged) fermion density is normalized to 1 for each level,
\begin{eqnarray}
1 &=& \frac{1}{a} \int_0^a {\rm d} x \, \psi^{\dagger} \psi
\\
& = &  |{\cal N}|^2\frac{1}{\ell}  \int_0^{\ell}{\rm d}\xi 
\left( |\phi_+|^2 + \frac{1}{\kappa^2\omega^2} |(\partial_{\xi} + \tilde{S})\phi_+ |^2 \right) \ . 
\nonumber
\label{s6}
\end{eqnarray}
The 2nd term can be simplified by partial integration and use of  the 2nd order
wave equation, Eq.~(\ref{7}). Due to the Bloch theorem the boundary terms vanish. 
In this way, one finds
\begin{equation}
1 = 2 |{\cal N}|^2 \frac{1}{\ell} \int_0^{\ell} {\rm d}\xi \, |\phi_+|^2 \ .
\label{s6a}
\end{equation}
We now insert the solution $\phi_+$ taken from the literature in terms of Jacobi functions
\cite{WhittakerWatson,Feinberg,Lietal}
\begin{eqnarray}
\phi_+ & = & \frac{H(\xi + \alpha) }{\Theta(\xi)} {\rm e}^{-\xi Z(\alpha)}
\nonumber \\
& = & \frac{\vartheta_1(v+w,q)}{\vartheta_4(v,q)}{\rm e}^{-\xi Z(\alpha)}
\label{s7}
\end{eqnarray}
with 
\begin{equation}
v=\frac{\pi \xi}{2 \mathbf{K}} \ , \qquad w = \frac{\pi \alpha}{2 \mathbf{K}} \ , \qquad q = {\rm nome}(\kappa) \ .
\label{s8}
\end{equation}
For the upper band, $\alpha={\rm i}\eta$. There is a 2nd solution, $\phi_+^*$,
which will simply be accounted for by a factor of 2 below.
For the definitions of the various Jacobi functions, see \cite{Jacobi}.
Using the following addition theorem for $\vartheta_1$ \cite{Wolfram}
\begin{equation}
\vartheta_3^2(0) \vartheta_1(x+y)\vartheta_1(x-y)=\vartheta_4^2(x)\vartheta_2^2(y)-\vartheta_2^2(x)\vartheta_4^2(y)
\label{s8a}
\end{equation}
together with standard
relations between different Jacobi functions \cite{Jacobi}, we find
\begin{equation}
|\phi_+|^2 = {\cal A}\left( 1- \frac{{\rm cn}^2(\xi|\kappa^2)}{{\rm cn}^2(\alpha|\kappa^2)}\right) 
\label{s9}
\end{equation}
with
\begin{equation}
{\cal A} =  \frac{\vartheta_2^2(w,q)}{\vartheta_3^2(0,q)} \ .
\label{s10}
\end{equation}
Now the ${\rm d}\xi$ integration in Eq.~(\ref{s6a}) can be performed 
(using the same variable transformation as in Eq.~(\ref{22}) as well as the relation $\ell = 2 \mathbf{K}$)
with the result
\begin{equation}
1=\frac{2 |{\cal N}|^2 {\cal A}}{\kappa^2 {\rm cn}^2(\alpha |\kappa^2)} \left( {\rm dn}^2(\alpha |\kappa^2)
-\frac{\mathbf{E}}{\mathbf{K}}\right) \ .
\label{s11}
\end{equation}
This determines the normalization factor $|{\cal N}|^2$.
Let us now consider the scalar density. In our representation of the Dirac matrices, Eq.~(\ref{2}), 
\begin{eqnarray}
\bar{\psi}\psi &=& - |{\cal N}|^2 (\phi_+^* \phi_- + \phi_-^* \phi_+) \nonumber \\
& = & \frac{|{\cal N}|^2}{\kappa \omega} (\partial_{\xi}+ 2 \tilde{S})|\phi_+|^2 \ .
\label{s12}
\end{eqnarray}
In the 2nd line, we have used Eq.~(\ref{s5}). Inserting the expression Eq.~(\ref{s9}) and performing
some straightforward calculations yields
\begin{equation}
\bar{\psi}\psi = \frac{2 {\cal A}|{\cal N}|^2}{\kappa^3 \omega}   \frac{{\rm dn}^2 (\alpha|\kappa^2)}
{{\rm cn}^2 (\alpha | \kappa^2)} \tilde{S} \ .
\label{s13}
\end{equation}
With the normalization factor $|{\cal N}|^2$ determined above and expressing $\tilde{S}$ through $S(x)$,
\begin{equation}
\bar{\psi}\psi = \frac{1}{\omega} \frac{\rm dn^2 (\alpha |\kappa^2)}
{\rm dn^2 (\alpha |\kappa^2)-\mathbf{E}/\mathbf{K}} S(x) \ .
\label{s14}
\end{equation}
This shows that the $x$-dependence of $\bar{\psi}\psi$ is the same for each orbit, only 
the prefactors differ.
In view of the relations [see \cite{Lietal} and Eq.~(\ref{12})]
\begin{eqnarray}
{\rm dn}^2 (\alpha |\kappa^2) &=&  {\cal E}-\kappa^2 \, ,\nonumber \\
\kappa \omega &=& \pm \sqrt{{\cal E} - \kappa^2}\, ,
\label{s15}
\end{eqnarray}
we get, for negative energy states, 
\begin{equation}
\bar{\psi}\psi = - \kappa  \frac{\sqrt{{\cal E}-\kappa^2}}
{{\cal E}-\kappa^2 - \mathbf{E}/\mathbf{K}} S(x) \ .
\label{s16}
\end{equation}
Finally we sum over all filled states. As in the calculation of the ground state energy,
we convert the integration over crystal momenta into an integration over ${\cal E}$,
include a factor of 2 for the twofold degeneracy of the orbits and employ the integration limits
Eqs. (\ref{18}-\ref{19}),
\begin{eqnarray}
\sum_{\alpha}^{\rm occ} \bar{\psi}_{\alpha} \psi_{\alpha} & = & 
\frac{2}{\kappa} \int_{{\cal E}_{\rm min}}^{{\cal E}_{\rm max}} \frac{{\rm d}{\cal E}}{2\pi}
\left| \frac{{\rm d}k}{{\rm d}{\cal E}}\right| \bar{\psi}\psi
\nonumber \\
&=&
-\frac{1}{2\pi}S(x)  \int_{{\cal E}_{\rm min}}^{{\cal E}_{\rm max}} {\rm d} {\cal E}
\frac{1}{\sqrt{(1-{\cal E})(1+ \kappa^2-{\cal E})}} \nonumber \\
& = & - \frac{1}{\pi} \ln ( \Lambda ) S(x)\ .
\label{s17}
\end{eqnarray}
We have dropped terms of order $1/\Lambda^2$ and higher, but of course no finite terms.
Inserting the relation between coupling constant and cutoff from the gap equation,
Eq.~(\ref{23}), then reproduces Eq.~(\ref{s1}) and proves the self-consistency of the scalar potential
(\ref{26b}).

\end{document}